\documentstyle[12pt]{article}
\begin{document}
THE QUANTUM COLLAPSE AND THE BIRTH \\OF A NEW UNIVERSE
\bigskip

M.L. Fil'chenkov\\
Alexander Friedmann Laboratory for Theoretical Physics,\\
26-9 Konstantinov Street, Moscow 129278, Russia\\
E-mail: fil@agmar.ru

\begin{abstract}
The gravitational collapse and the birth of a new universe are considered in
terms of quantum mechanics. Transitions from annihila\-tion of matter to
deflation in the collapse and from inflation to creation of matter in the
birth of a universe are considered. The creation of a new universe takes
place in another space-time since beyond the event horizon the time coordinate
is inextensible for an external observer. A reasonable probability of this
creation is obtainable only for miniholes. The major part of the mass of such
collapsing compact objects as stars, quasars and active nuclei of galaxies
remains confined in the potential well near the vacuum state.
\end{abstract}

\section{Introduction}
The gravitational collapse problem is closely related to the cosmological
problem. This has first been revealed in\quad\cite{HTWW} where the geometry
of a collapsing body in the comoving coordinates was proved to be a geometry
of Friedmann's closed world. At the same time it is a quantum consideration of
both the late collapse and the early Universe that is required for solving
the above-mentioned problems.

It is also of interest to consider quantum-mechanically a birth of the
universe in the laboratory\quad\cite{FG, N}.  Some people consider a classical
transition from the collapse to the birth of a new universe\quad\cite{FMM, BF}
and an eternal nonsingular black holes as a final stage of the collapse\quad
\cite{D}.

In the present paper we use the approach introduced in\quad\cite{FPLB, FAAT}.
The collapse in our universe is assumed to give rise to the birth of a new
one. From the viewpoint of quantum mechanics this is a tunnelling. In our
space-time the collapse ends in approaching Schwarzschild's horizon, i.e its
final stage is unobservable. The new universe is being born in another space-
time.

\section{Approach}
In case the scale factor $a$ and the scalar field $\phi$ are chosen as
independent variables, Wheeler-DeWitt's equation in minisuperspace has the
form\quad\cite{L}
$$
[-\frac{1}{a^p}\frac{\partial}{\partial{a}}a^p\frac{\partial}{\partial{a}}
+\frac{1}{a^2}\frac{{\partial}^2}{\partial{\phi^2}}+V(a,\phi)]\psi=0 \eqno(1)
$$
where $p$ takes account of the operator ordering (below we assume $p=0$).\\
Comparing (1) with Klein-Gordon's equation
$$
-c^{2}\triangle\psi+\frac{{\partial}^{2}\psi}{\partial t^2}+\frac{m^{2}
c^{4}}{\hbar^2}\psi=0,                                        \eqno(2)
$$
it is easy to see that in the Lorentzian domain ($V<0$) $a$ plays the role of
time and $\int_{}^{}a\,d\phi$ -- the role of a coordinate, whereas in the
Euclidean domain ($V>0$) $\int_{}^{}a\,d\phi$ plays the role of time and $a$
-- the role of a coordinate. For $\frac{\partial\psi}{\partial\phi}=0$ (1)
reduces to
$$
-\frac{d^2\psi}{da^2}+V(a)\psi=0.                                \eqno(3)
$$

The independence from $\phi$ is equivalent to homogeneity in the Lorentzian
domain and stationarity in the Euclidean one. In Friedmann's universe the
potential
$$
V=\frac{2m_{pl}}{\hbar^2}[U(a)-E]                                  \eqno(4)
$$
allows one to reduce (3) to Schr\"odinger's equation\quad\cite{FPLB,FAAT},
where
$$
U(\gamma)=\frac{m_{pl}c^2}{2}(k\gamma^{2}-B_{0}\gamma^{4}-B_{1}\gamma^{3}-
B_{2}\gamma^{2}-B_{3}\gamma-B_{5}\gamma^{-1}-B_{6}\gamma^{-2}),    \eqno(5)
$$
$k$ is the model parameter, $E=\frac{1}{2}m_{pl}c^{2}B_{4}$, where $\gamma=
\frac{a}{l_{pl}}$, $B_{n}$ are the coefficients of the expansion of the energy
density in Laurent's series in $\gamma$:
$$
\varepsilon=\varepsilon_{pl}\sum_{n=0}^6 B_{n}\gamma^{-n},         \eqno(6)
$$
$$
n=3(1+\alpha),                                                     \eqno(7)
$$
where $\alpha$ is a coefficient in the equation of state $p=\alpha\varepsilon
$ ($n=0$ vacuum, $n=1$ domain walls, $n=2$ strings, $n=3$ dust, $n=4$
ultrarelativistic gas, $n=5$ perfect gas, $n=6$ ultrastiff matter).

Eq. (3) and formulae (4)--(7) describe the universe behaving as a planckeon
with the energy of an ultrarelativistic gas $E=\frac{1}{2}m_{pl}c^{2}B_{4}$
moving in the field of the rest matter ($B_{n}\ne B_{4}$).

\section{Quantum Collapse}

Since the geometry of a collapsing body in the comoving coordinates mimics
Friedmann's closed model, the form of the potential in Schr\"odinger's
equation should be the same. However, there arises some difference due to the
 equation of state of the collapsing body undergoing a sudden change from
$p=\frac{\varepsilon}{3}$, $p=\frac{2}{3}\varepsilon$, $p=\varepsilon$
(annihilation of matter\footnote{by annihilation of matter we mean not a
process inverse to pair creation but that related to nonconservation of a
baryon charge})to $p=-\varepsilon$ (deflation). This can be taken into
account in modifying (5) by adding the terms with negative powers of
$\gamma-\gamma _{0}$ , where $\gamma_{0}$ corresponds to the scale factor  at
which the equation of state undergoes the change. Finally the potential
takes the form
$$
U(\gamma)=\frac{m_{pl}c^{2}}{2}[k\gamma^{2}-B_{0}\gamma^{4}-B_{1}\gamma^{3}-
B_{2}\gamma^{2}-B_{3}\gamma-B_{5}\gamma^{-1}-B_{6}\gamma^{-2}
$$
$$
-B'_{5}(\gamma-\gamma_{0})^{-1}-B'_{6}(\gamma-\gamma_{0})^{-2}],    \eqno(8)
$$
where $k=+1$.

For $\gamma\gg\sqrt[4]{\frac{2E}{m_{pl}c^2}}\gg1$ and $B_{0}=1$ Eq. (3)
reduces to
$$
\frac{d^2\psi}{d\gamma^2}+\gamma^4\psi=0,                          \eqno(9)
$$
which has a solution\quad\cite{K}
$$
\psi=\gamma^{\frac{1}{2}}Z_{\frac{1}{6}}(\frac{\gamma^3}{3}).     \eqno(10)
$$
with the asymptotics \quad\cite{L}
$$
\psi=C_1e^{i\frac{\gamma^3}{3}}+C_2e^{-i\frac{\gamma^3}{3}}.       \eqno(11)
$$
This WKB formula corresponds to the classical inflation $\gamma=e^{t/t_{pl}}$
or deflation $\gamma=e^{-t/t_{pl}}$ since the action $S\propto(\pm a^3)$ and
on the other hand $S=\int_{}^{}L\,d\eta$, where the Lagrangian $L\propto a^4$
and $d\eta=dt/a$ ($\eta$ is the conformal time), hence $\dot a=\pm Ha$ and
$a=Ce^{\pm Ht}$ with $H=1/t_{pl}$.

For $\gamma-\gamma_{0}\ll1$ ($B_{0}=1$) Eq. (3) takes the form
$$
\frac{d^2\psi}{d\gamma^2}+[\gamma_{0}{}^4+B'_{5}(\gamma-\gamma_{0})^{-1}+
B'_{6}(\gamma-\gamma_{0})^{-2}+\frac{2E}{m_{pl}c^2}]\psi=0,         \eqno(12)
$$
which has a solution\quad\cite{LL}
$$
\psi={\rm const}\times\rho^{s+1}e^{-\frac{\rho}{2}}F(-p, 2s+2, \rho),\eqno(13)
$$
satisfying the boundary condition $\psi(\gamma_0)=0$,\\
where $F$ is a degenerate hypergeometric function,
$$
\rho=2(\gamma-\gamma_{0})\sqrt{-\gamma_{0}{}^4-\frac{2E}{m_{pl}c^2}},
$$
$$
n=\frac{B'_{5}}{2\sqrt{-\gamma_{0}{}^4-\frac{2E}{m_{pl}c^2}}}\quad (B'_{5}>0),
$$
$$
s=-\frac{1}{2}+\sqrt{\frac{1}{4}-B'_{6}},
$$
$$
 n-s-1=p=0, 1, 2,...
$$
The energy spectrum has the form
$$
E_{p}=-\frac{B'_{5}{}^2m_{pl}c^2}{8(p+\frac{1}{2}+
\sqrt{\frac{1}{4}-B'_{6}})^2}-\frac{{\gamma_{0}}^4}{2}m_{pl}c^2.     \eqno(14)
$$

It should be noted that this discrete spectrum does exist only if the
conditions
$$
\gamma_{0}{}^4\le\frac{2|E|}{m_{pl}c^2}\le\frac{2M}{m_{pl}}         \eqno(15)
$$
and
$$
B'_{6}\le\frac{1}{4}                                                \eqno(16)
$$
are valid, where $M$ is the rest mass of a collapsing body. As seen from (8),
(15) results in the absolute value of the vacuum energy not exceeding
$|E|$. The last equality in (15) is obtainable if one assumes $E+Mc^2=m_{pl}
c^2$, where $M\gg m_{pl}$.

The  compact objects are known to collapse for
$M=10-10^{9}M_{\odot}$: stars at $M\sim10M_{\odot}$ and quasars and active
nuclei of galaxies at $M\sim10^{9}M_{\odot}$. Hence
$$
\gamma_{0}\le\sqrt[4]{\frac{2M}{m_{pl}}}=\sqrt[4]{\frac{r_{g}}{l_{pl}}},
$$
i.e. $\gamma_{0}\le10^{10}$ or $a_{0}\le10^{-23}$cm for stars and $\gamma_{0}
\le10^{12}$ or $a_{0}\le10^{-21}$cm for quasars and active nuclei of galaxies.

Eq. (12) reduces to Eq. (9) at $\gamma=\gamma_{0}+1$ for $\gamma_{0}\gg1$
if $B'_{5}+B'_{6}+B_{4}=0$, which results in $B'_{5}\approx=-B_{4}$ since
$B'_{6}\le\frac{1}{4}\ll|B_{4}|$ ($B'_{6}\ge0$). Assuming $B'_{5}=-B_{4}$,
$B'_{6}=0$, we obtain
$$
B_{4}=-2n^2+2n\sqrt{n^2-\gamma_{0}{}^4}                            \eqno(17)
$$
where
$B_{4}=-2\gamma_{0}{}^4$ at $n=\gamma_{0}{}^2$ and $B_{4}=-\gamma_{0}{}^4$ for
$n\gg\gamma_{0}{}^2$. This means that only higher levels are possible for
$\gamma_{0}\gg1$.

From (6) it follows that the absolute value of the partial
energy density $|\varepsilon_4|=\varepsilon_{pl}\frac{B_4}{\gamma_0{}^4}\sim
\varepsilon_{pl}$ at $\gamma=\gamma_0$ since $\frac{|B_4|}{\gamma_0{}^4}\sim1
$.

For $\gamma$ very close to $\gamma_{0}$, when $\gamma-\gamma_{0}\ll\sqrt{\frac
{B'_{6}m_{pl}c^2}{2E}}$, Eq. (12) reduces to
$$
\frac{d^{2}\psi}{d\gamma^2}+\frac{B'_{6}}{(\gamma-\gamma_{0})^
{2}}\psi=0,                                                        \eqno(18)
$$
which has a solution\quad\cite{K}
$$
\psi=\sqrt{\gamma-\gamma_{0}}\left\{\begin{array}{lcr}
C_{1}\cos[(b\ln(\gamma-\gamma_{0})]+C_{2}\sin[b\ln(\gamma-\gamma_{0})],
\quad b^{2}=B'_{6}-\frac{1}{4}>0;\\
C_{1}(\gamma-\gamma_{0})^{b}+C_{2}(\gamma-\gamma_{0})^{-b}, \quad{b}^2=\frac
{1}{4}-B'_{6}>0;\\
C_{1}+C_{2}\ln(\gamma-\gamma_{0}), \quad B'_{6}=\frac{1}{4}.\\
\end{array}\right.                                                 \eqno(19)
$$
For $B'_{6}$ there occurs a fall to the field centre, which corresponds to
$E_{0}=\infty$\quad\cite{LL}.

\section{Birth of a Universe}

After the tunnelling through a potential barrier the collapsing body appears
in the potential well where for $\gamma\ll1$ Schr\"odinger's equation reduces
to\quad\cite{FPLB, FAAT}
$$
\frac{d^{2}\psi}{d\gamma^2}+(\frac{B_{5}}{\gamma}+\frac{B_{6}}{\gamma^2}+
\frac{2E}{m_{pl}c^2})\psi=0                                       \eqno(20)
$$
whose solution is given by formula (13), satisfying the boundary condition
$\psi(0)=0$,\\
where
$$
\rho=2\gamma\sqrt{\frac{-2E}{m_{pl}c^2}},
$$
$$
n=\frac{B_{5}}{2\sqrt{\frac{-2E}{m_{pl}c^2}}}.
$$
The energy spectrum has the form
$$
E_{p}=-\frac{B_{5}{}^2m_{pl}c^2}{8(p+\frac{1}{2}+\sqrt{\frac{1}{4}-
B_{6}})}.                                                         \eqno(21)
$$
For very small $\gamma$, when $\gamma\ll\sqrt{\frac{B_{6}m_{pl}}{2E}}$, $\psi$
-function has the form (19) with $\gamma-\gamma_{0}$ substituted by $\gamma$
and $B'_{6}$ by $B_{6}$.
The WKB coefficient for penetration through the potential barrier at $a>0$
reads\quad\cite{LL}
$$
D=\exp[-\frac{2l_{pl}}{\hbar}|\int_{\gamma_{1}}^{\gamma_{2}} \sqrt{2m_{pl}
(E-U)}\,d\gamma|]                                                \eqno(22)
$$
where $\gamma_{1}\approx\sqrt[4]{-B_{4}}$, $\gamma_{2}\approx-\frac{B_{5}}
{B_{4}}$, $E=\frac{1}{2}m_{pl}c^2B_{4}$. U is given by formula (8) with
$B'_{5}=B'_{6}=0$ since we consider tunnelling in the domain where
$\gamma<\gamma_{0}$. For $|B_{4}|\gg1$ we obtain
$$
D\approx e^{-2\sqrt[4]{(-B_{4})^3}}.                                \eqno(23)
$$
According to (14) $|B_{4}|\le\frac{2M}{m_{pl}}$. Hence we obtain
$D\ge e^{-10^{29}}$ for stars, $D\ge e^{-10^{35}}$ for quasars and active
nuclei of galaxies.\\
The penetration factor given by formula (22) has first been calculated by
G. Gamow \quad\cite{G} for the case of alpha decay in radioactive nuclei.
Gamow's procedure was extended in \quad\cite{KM,AP,Z,GZ,V,HH,L,DZS} to the
case of the birth of the Universe from a pure vacuum.

The probability of the birth of a new universe, as a result of the
gravitational collapse in our space, is $W=D^2$ due to penetration through
two barriers: at $a>0$ in our space-time and at $a<0$ in another space-time
where the birth takes place. For the sake of simplicity, we may identify the
other space-time with the negative semiaxis of the scale factor. The universe
cannot be born in the space-time where the gravitational collapse occurs,
because for an external observer the time coordinate is inextensible beyond
the event horizon, i.e. through the point with $t=+\infty$.
Then the scenario in the new universe (other space-time) might be a mirror
reflexion of that in our space-time. The total scenario for $-\infty<\gamma<+
\infty$ is determined by the potential $U(|\gamma|)$. Near the singularity at
 $\gamma=0$ there will be twice as many energetic levels as given by
formula (14). For the new universe at $\gamma\approx-\gamma_{0}$ there occurs
a transition from the equation of state $p=-\varepsilon$ (inflation) to $p=
\frac{\varepsilon}{3}$, $p=\frac{2}{3}\varepsilon$, $p=\varepsilon$ (creation
of matter). The same transition takes place in our universe at $\gamma\approx
\gamma_{0}$ (reheating after inflation \quad\cite{DZS}).

To create our universe with the mass $M=10^{55}$g\quad\cite
{L} (if it is assumed to be closed), it is neccesary for the parameter $\gamma
_{0}=e^{t/t_{pl}}$ to be equal to $10^{15}$. Hence the time of the end of
inflation is $t=35t_{pl}$, which agrees with classical estimates within a
factor of two\quad\cite{L, DZS}. The WKB penetration factor $D\ge e^{-10^{45}
}$ for our universe, the probability of its creation $W\ge e^{-2\cdot10^{45}}
$.

The infintesimal values of the penetration factor for collapsing stars,
galaxies, quasars or for the birth of our universe mean that  not all
mass of these objects, may be a small part of it, tunnels. The major part of
the mass of these objects is confined in the potential well near $\gamma=
\gamma_0$, whose left boundary corresponds to the vacuum state,  never
reaching the singularity at $\gamma=0$.

If, e.g., we assume that $\gamma_0=10$, $B_4=-10^4$ and $M=5\cdot10^3$, then
$D=e^{-2\cdot10^3}$. Thus, a reasonable penetration factor is obtainable only
for miniholes. As to Hawking's effect, it is easy to see that the evaporation
time $\tau_H=\frac{Mc^2}{P_H}$, where $P_H=\frac{\hbar\kappa^2}{960\pi c^2}$
\quad \cite{FA}, reads
$$
\tau_H=1.536\cdot10^4(\frac{M}{m_{pl}})^3t_{pl}.                \eqno(24)
$$
Whereas the deflation time
$$
t_d=\frac{t_{pl}}{4}\ln\frac{2M}{m_{pl}}.                       \eqno(25)
$$
Thus $t_d\ll\tau_H$ for $M>m_{pl}$. Hence Hawking's effect is negligible as
compared with the deflation. Apropos, from (25) it follows that the minimum
collapsing mass $M=\frac{1}{2}m_{pl}$.

\section{Conclusion}

We have considered the total cycle from the quantum gravitational collapse of
a body to the birth of a new universe in another space-time. However, the
tunnelling that accompanies the collapse occurs with an extremely low
probability, which makes us assume that not all mass of the collapsing compact
object, such as a star or a galaxy, tunnels through the barriers separating it
from the singularity or another space-time where a new universe emerges. The
major part of the mass of these objects remains confined in the potential well
near the vacuum state. The tunnelling giving rise to a new universe may occur
only for miniholes.

\section{Acknowlegement}

I am grateful to A.A. Grib, I.G. Dymnikova, G. Esposito, B. Dragovich,
R.X. Saibatalov, E.I. Guendelman and Yu.V. Grats for useful discussions.


\begin{thebibliography}{99}
\bibitem{HTWW} B.K. Harrison, K.S. Thorne, M. Wakano, J.A. Wheeler,
Gravitational Theory and Gravitational Collapse (Univ. Chicago Press, Chicago
and London, 1965).
\bibitem{FG} E.Farhi, A.H. Guth, Phys. Lett. B 183 (1987) 149.
\bibitem{N} I.D. Novikov, How the Universe Exploded (Nauka, Moscow, 1988).
\bibitem{FMM} V.P. Frolov, M.A. Markov, V.F. Mukhanov, Phys. Rev. D 41 (1990)
383.
\bibitem{BF} C. Barrab\`es, V.P. Frolov, Phys. Rev. D 53 (1996) 3215.
\bibitem{D} I. Dymnikova, GRG, 24 (1992) 235.
\bibitem{FPLB} M.L. Fil'chenkov, Phys. Lett. B 354 (1995) 208.
\bibitem{FAAT} M.L. Fil'chenkov, Astron. \& Ap. Trans. 10 (1996) 129.
\bibitem{L} A.D. Linde, Elementary Particle Physics and Inflationary
Cosmology (Nauka, Moscow, 1990).
\bibitem{K} E. Kamke, Differentialgleichungen. L\"osungsmethoden und
L\"osungen. I. Gew\"ohnliche Differetialgleichungen (Leipzig, 1959).
\bibitem{LL} L.D. Landau, E.M. Lifshitz, Quantum Mechanics. Nonrelativistic
Theory (Nauka, Moscow, 1963).
\bibitem{G} G. Gamow, Z. Phys. 51, 3-4 (1928) 204.
\bibitem{KM} M.I. Kalinin, V.N. Melnikov, Trud. VNIIFTRI 16(46) (1972) 43.
\bibitem{AP} D. Atkatz, H. Pagels, Phys Rev. D 25 (1982) 2065.
\bibitem{Z} Ya.B. Zeldovich, Pis'ma Astr. Zhurn. 85 (1981) 209.
\bibitem{GZ} L.P. Grishchuk, Ya.B. Zeldovich, In: Quantum Structure of
Space-Time, Eds. M. Duff, C. Isham (Cambridge Univ. Press, Cambridge, 1983),
p. 353.
\bibitem{V} A. Vilenkin, Phys Rev. D 30 (1984) 509; Nucl. Phys. B 252 (1985)
141; Phys. Rev. D 50 (1994) 2581.
\bibitem{HH} J.B. Hartle, S.W. Hawking, Phys. Rev. D 28 (1983) 2860.
\bibitem{DZS} A.D. Dolgov, Ya.B. Zel'dovich, M.V. Sazhin, Cosmology of the
Early Universe (Moscow Univ. Press, Moscow, 1988).
\bibitem{FA} M.L. Fil'chenkov, GR15, Abstr. Plen. Lect. Contr. Papers, Pune.-
1997, p. 279.
\end{thebibliography}
\end{document}